# Emergent of the flat band and superstructures in the VSe$_2$ / Bi$_2$Se$_3$ system

Turgut Yilmaz[1]*, Xiao Tong[2], Zhongwei Dai[2], Jerzy T. Sadowski[2], Eike F. Schwier[3], Kenya Shimada[3], Sooyeon Hwang[2], Kim Kisslinger[2], Konstantine Kaznatcheev[1], Elio Vescovo[1], and Boris Sinkovic[4]

[1]National Synchrotron Light Source II, Brookhaven National Lab, Upton, New York 11973, USA

[2]Center for Functional Nanomaterials, Brookhaven National Lab, Upton, New York 11973, USA

[3] Hiroshima Synchrotron Radiation Center, Hiroshima University, 2-313 Kagamiyama, Higashi Hiroshima 739-0046, Japan

[4] Department of Physics, University of Connecticut, Storrs, Connecticut 06269, USA

*tyilmaz@bnl.gov

**Dispersionless flat bands are proposed to be a fundamental ingredient to achieve the various sought after quantum states of matter including high-temperature superconductivity[1-4] and fractional quantum Hall effect[5-6]. Materials with such peculiar electronic states, however, are very rare and often exhibit very complex band structures. Here, we report on the emergence of a flat band with a possible insulating ground state in the sub-monolayer VSe$_2$ / Bi$_2$Se$_3$ heterostructure by means of angle-resolved photoemission spectroscopy and scanning tunneling microscopy. The flat band is dispersionless along the k$_{II}$ and k$_z$ momenta, filling the entire Brillouin zone, and it exhibits a complex circular dichroism signal reversing the sign at several points of the Brillouin zone. These properties together with the presence of a Moiré patterns in VSe$_2$ suggest that the flat band is not a trivial disorder or confinement effect and could even be topologically non-trivial. Another intriguing finding is that the flat band does not modify the Dirac cone of Bi$_2$Se$_3$ around the Dirac point. Furthermore, we found that the flat band and the Dirac surface states of Bi$_2$Se$_3$ have opposite energy shifts**

**with electron doping. This opens a novel way of controlling the spin texture of photocurrents as well as the transport properties of the heterostructure. These features make this flat band remarkably distinguishable from previous findings and our methodology can be applied to other systems opening a promising pathway to realize strongly correlated quantum effects in topological materials.**

The physics of solids is determined by their energy band structures. Therefore, investigation and controlling distinct electronic band dispersions are of great importance in condensed matter to understand and discover new states of the matter. One of the exotic electronic states is a type of flat band which is predicted to host high-temperature superconductivity[1-4], fractional quantum Hall effect (FQHE)[5-6], and ferromagnetism[7-9]. In a superconductor, a flat band can boost the coupling constant and the transition temperature ($T_c$) because of the enhanced density of states at the Fermi level ($E_F$)[10-11]. This was utilized to explain the unexpected superconductivity in rhombohedral graphite and twisted graphene[12-14]. Other examples of the flat band materials are Kagome lattices in which the flat band stems from the destructive quantum interference due to the frustrated lattice geometry[7-8]. The Kagome-type flat band was reported in FeSn and $Fe_3Sn_2$ by angle-resolved photoemission spectroscopy (ARPES) and in $Co_3Sn_2S_2$ by scanning tunneling spectroscopy (STS)[8-10]. However, the complexity of the electronic structure in photoemission data and the lack of momentum resolution in STS make the observations elusive. Another system with the dispersionless flat electronic states could be the Mott insulators in which the strong correlation effect forms the flat Hubbard band with an insulating gap the $E_F$ as predicted in $TaS_{2-x}Se_x$ and $NbS_{2-x}Se_x$[15-18]. The strong correlation effect was also investigated in $Bi_2Se_3$ and $Bi_2Te_2Se$ topological insulators (TI) and a possible p-band Mott insulating state where the Hubbard bands were predicted to exist in these samples[19-21]. The experimental signature of the Mott insulating state was also captured in multilayer graphene Moiré superlattice indicating the strong correlation effect in that system[22]. Computational efforts were also focused on designing the flat band transition metal dichalcogenides (TMDs) Moiré superlattices which could support strongly correlated physics at higher-temperatures due to the flat bands[23-25]. Thereby, the previous works conclude that the flat band media could be fertile to many novel states of the matter. However, apart from Mott insulators, the limited number of the materials with such non-trivial bands hinders future studies.

Motivated by earlier studies, we investigate the surface electronic structure of $VSe_2$ TMD grown on the surface of $Bi_2Se_3$ TI and show the emergence of a flat band in the electronic states at low $VSe_2$ coverage level. This flat band covers the entire $k_x$ - $k_y$ plane of the Brillouin zone (BZ) and it displays dispersionless along the $k_z$ direction as well. Furthermore, our circular dichroism ARPES (CD-ARPES) measurements revealed that the CD signal of the flat band reverses the sign at several points within the BZ. Another notable observation is that $VSe_2$ growth and emergence of the flat band do not reshape the Dirac cone of $Bi_2Se_3$ in the vicinity of the DP unlike the case of transition metal doping which opens a large gap at the DP[26]. We also observed Moiré patterns in $VSe_2$ domains of monolayer (ML) thickness and stripe type patterns in bare $Bi_2Se_3$ through scanning tunneling microscopy (STM). Along with these observations, electron doping impact on the band structure of the system suggests that the flat band might be due to the Mott insulator like interaction correlated with the charge density wave (CDW) phase. To further shed light on the crystalline and chemical properties of the system, scanning transmission electron microscopy (STEM), micro-spot low energy electron diffraction (μLEED), and x-ray photoemission spectroscopy (XPS) experiments were conducted. Our results demonstrate rich physics and suggest a large family of materials of possible emergent flat bands and thus will motivate future studies on materials with superconductivity at high-temperatures, QSL states, or FQHE by using our approach.

**Structural characterization:** $Bi_2Se_3$ and $VSe_2$ crystals are layered materials with their atomic stacking geometry shown in Fig. 1a. The layers in each compound are separated by so-called van der Waals (vdW) gaps with weak covalent out-of-plane bonds connecting the layers. This is the reason that $VSe_2$ / $Bi_2Se_3$ heterostructure can be grown despite the large in-plane lattice mismatch of around 20 % between the two materials[27-28]. Figs. 1b and 1c depict the relevant core-levels of such structures formed by the growth of 0.3 ML and 3 ML $VSe_2$ on 12 quintuple layer (QL) $Bi_2Se_3$. Upon deposition of the $VSe_2$, the spectral shape of the Bi 5d peaks of $Bi_2Se_3$ does not exhibit a prominent modification indicating the absence of V metals at the interface and/or in the bulk (Fig. 1b)[29]. Compared with pristine $Bi_2Se_3$, the Se 3d peak, however, appears at 0.1 eV higher binding energy upon $VSe_2$ deposition. This can be better seen in the inset of Fig. 1b displaying the scaled Se 3d peaks to the same height. The difference in binding energy could be due to the CDW phase of $VSe_2$[27]. V $2p_{1/2}$ and $2p_{3/2}$ peaks shown in Fig. 1c are located at 513 eV and 520.6 eV binding energies, being in agreement with recent reports[28].

To further explore the system, we show a high-angle annular dark-field (HAADF)-STEM cross-section image of 3 ML $VSe_2$ / 12 QL $Bi_2Se_3$ heterostructure in Fig. 1d. $Bi_2Se_3$ and $VSe_2$ exhibit regular atomic layers with smooth interfaces and vdW gaps marked with red arrows in Fig. 1d. On the other hand, the STEM data does not exhibit a clear interface spacing between the $VSe_2$ and $Bi_2Se_3$ which could strongly modify the local electronic structure. Furthermore, the STEM energy dispersive X-ray spectroscopy elemental maps presented in Fig. S1 (supplementary materials) shows the atomic distribution of Bi in the $Bi_2Se_3$ layers, V in $VSe_2$ layers, and Se across the heterostructure as expected.

Fig. 1e and 1f show the STM image of $Bi_2Se_3$ and $VSe_2$ regions of a 0.3 ML $VSe_2$ / 12 QL $Bi_2Se_3$ sample, respectively. The $Bi_2Se_3$ surface has a stripe-like pattern similar to Cs and Fe doped $Bi_2Se_3$[30]. The STM image of the $VSe_2$ domains presented in Fig. 1f exhibits a Moiré pattern with ~ 2 nm x 2 nm superstructure. This differs from the previous studies conducted on $VSe_2$ / graphene in which such superstructure formation was absent[27]. Moiré pattern can be formed by a small misfit between the in-plane lattice parameters of the film and the underlying material or the relative rotation of two layers to each other, or both. By contrast, the lattice mismatch between the $VSe_2$ and $Bi_2Se_3$ is quite large (about 20 %). Unfortunately, we cannot make a quantitative analysis for precise determination of the in plane lattice parameters or the atomic displacement due to the limitation in our STM data taken at room temperature experiment. But, similar Moiré pattern formation is also observed on monolayer $MoSe_2$ grown on a graphene substrate whose origin is attributed to the lattice mismatch between the multiple unit cells of the two materials[31]. Thereby, the Moiré pattern in $VSe_2$ could be formed due to the small mismatch between four-five-unit cells of $Bi_2Se_3$ ($4a_{BS}$ = 16.56 Å or $5a_{BS}$ = 20.7 Å) and five-unit cells of $VSe_2$ ($5a_{VS}$=16.8 Å or $6a_{VS}$=20.16 Å) for the rotationally aligned lattice geometry. Alternatively, the Moiré pattern could form be formed by the rotational misalignments of $Bi_2Se_3$ and $VSe_2$ atomic lattices. Moreover, the details of the STM data reveal that the layer heights are 6.8 Å for $VSe_2$ and 9.6 Å $Bi_2Se_3$ (Fig. S2, supplementary materials), which is in agreement with published results[32].

**ARPES electronic structure:** To examine the band structure, the binding energy versus $k_y$ plots are given for 12 QL $Bi_2Se_3$ in Fig. 2a and for 0.3 ML $VSe_2$ / 12 QL $Bi_2Se_3$ in Fig. 2b. The $Bi_2Se_3$ sample exhibits the typical band structure with the linear Dirac surface states (DSSs) forming the Dirac cone with the Dirac point (DP) at 0.36 eV below $E_F$[33]. Upon deposition of 0.3 ML $VSe_2$ on $Bi_2Se_3$, a flat band with 0.47 eV binding energy and ~0.18 eV bandwidth emerges in the surface electronic structure (Fig. 2b). The flatness of the band is well distinguished in the energy distribution curves (EDCs) shown in Fig. 2c. The bulk bands and the DSS of $Bi_2Se_3$ strongly disperse as a function of $k_y$, while the flat band retains dispersionless across the $\bar{\Gamma} - \bar{M}$ high symmetry lines. $VSe_2$ growth also induces the well-known M-state quantization[34] of the bulk valance band (BVB) of $Bi_2Se_3$ shown in Fig. 2b. Moreover, the flat band overlaps with the lower branch of the Dirac cone in the vicinity of $k_y = 0$ Å$^{-1}$ without inducing a prominent change in its spectral shape. This can be even better seen in the films with larger $VSe_2$ coverage confirming that the flat band, DSSs, and the dispersive V 3d state of $VSe_2$ coexist in the surface electronic structure (Fig. S3, supplementary materials). Also, ARPES date for the thicker $VSe_2$ coverage suggests that the flat band is localized on the bare $Bi_2Se_3$ and/or at the interface between $VSe_2$ and $Bi_2Se_3$ ( Fig. S4, supplementary materials). We should also note that 0.3 ML $VSe_2$ / 12 QL $Bi_2Se_3$ sample shows an insulating gap at the $E_F$ which will be discussed in the following sections of the text.

To further investigate the flat band, a $k_x - k_y$ intensity plot at $E_F$ and the binding energy of the flat band ($E_{FB}$) are shown in Fig. 2d and in Fig. 2e, respectively. The Femi surface is dominated by a circular counter of the DSSs centered at $\bar{\Gamma}$ point. The constant energy cut at $E_{FB}$ reveals another remarkable feature, namely that the flat band reaches beyond the $\bar{M}$ and $\bar{K}$ points in the BZ. This observation reveals that the flat band fills the entire BZs of $Bi_2Se_3$ and $VSe_2$ as depicted by black and red hexagons in Fig. 2e, respectively. Such electronic state spread over large momentum area can significantly enhance the electronic correlation yielding quantum effects at very high-temperatures. It is also worth noting that the LEED pattern of the sample shows rotationally stretched diffraction spots along the rotational direction indicating the presence of the rotationally misaligned $VSe_2$ domains ( Fig. S2a, supplementary materials) with respect to each other and to the $Bi_2Se_3$ substrate. The rotational misfit of ±3º estimated from µLEED pattern, however, is too small for a band to span whole BZ and to induce a fully occupied constant energy counter in the momentum map.

**Photon energy-dependent ARPES:** In ARPES experiments, changing the photon energy corresponds to mapping the electronic states along the $k_z$ direction of the BZ. By recording the electronic structure with a wide photon energy range, a $k_\parallel$ vs. $k_z$ or binding energy vs. $k_z$ dispersions can be also extracted. This method allows studying the energy bands along the $k_z$ (ou-of-plane) direction to distinguish the non-dispersive states from the dispersive bulk bands. Such spectra acquired at varying photon energies (40 eV to 70 eV) for 0.3 ML $VSe_2$ / 12 QL $Bi_2Se_3$ sample are given in Fig. 3. In the plot of $k_z$ versus $k_y$ dispersion at $E_F$, DSSs marked with dashed red lines exhibit no $k_z$ dependence (Fig. 3a). Similar spectrum at $E_{FB}$ given in Fig. 3b shows that a high spectral intensity along the $k_y = 0$ Å$^{-1}$ originates from the bottom of the Dirac cone of $Bi_2Se_3$. Away from the $k_y = 0$ Å$^{-1}$, the plot has non-vanishing spectral intensity contributed by the flat band. This can be better seen in the momentum distribution curves (MDCs) obtained at various $k_z$ points (Fig. 3c) in which each spectrum exhibits always finite density of states along the $k_y$ momentum direction. This implies the dispersionless nature of the flat band along the $k_z$ momentum direction. To further validate this observation, we show the binding energy - $k_z$ plots along the $k_y = \pm 0.3$ Å$^{-1}$ directions in Fig. 3d and 3e, respectively. The plots clearly show that the flat band at 0.47 eV binding energy is $k_z$ - independent confirming its non-bulk derived nature. We should also note that the M-shape bulk band located at 1 eV binding energy exhibits a nearly non-dispersive feature along the $k_z$ as shown in Fig. 3d and 3e. To further reveal the details of the flat band, Fig 3f depicts the EDCs taken at different $k_z$ points. One can see that the EDC of the flat band does not exhibit a $k_z$ dependent evolution in the binding energy and bandwidth providing a signature that the flat band originates from a single band.

**CD-ARPES:** Circular dichroism (CD)-ARPES has gained great attention due to its feasibility to investigate the helical spin-orbit texture in topological surface states[35]. The principle of the method is the spectral weight differences in ARPES arising from the opposite helicity of the circularly polarized light. CD-ARPES is then obtained from $[(I_{RCP} - I_{LCP})]/[(I_{RCP} + I_{LCP})]$ where $I_{RHP}$ and $I_{LHP}$ are photoemission intensities for RCP and LCP lights, respectively. Thus, we have recorded the band structure of 0.3 ML $VSe_2$ / 12 QL $Bi_2Se_3$ sample with RCP and LCP, shown in Fig. 4a and in Fig. 4b, respectively. The corresponding CD-ARPES is presented in Fig. 4c with red (negative-CD) and blue (positive-CD) color representations. The bulk bands of $Bi_2Se_3$ dispersing below 0.8 eV binding energy show a strong DC as seen in Fig. 4a-4c. CD from the DSSs of $Bi_2Se_3$ is also seen switching the spectral weight from the $-k_y$ to $+k_y$ regions when changing the excitation from RCP to LCP. For clarity, the DC signal versus $k_y$ is plotted in Fig. 4d at 0.1 eV binding energy in which the CD is positive for left and negative for the right side of the Dirac cone, marked with vertical arrows. Further away from the $k_y = 0$ Å$^{-1}$, the plot in Fig. 4d still shows non-zero CD. This is likely originating from the V 3d orbitals which dominate the $E_F$ density of states in $VSe_2$[27]. Also, since the spectra are conducted with 50 eV photons corresponding to nearly Z-point of the BZ, the BCBs of $Bi_2Se_3$ do not contribute to ARPES spectra in the vicinity of the $E_F$ as seen in Fig. 2a.

To investigate the dichroism effect in the flat band, the CD at $E_{FB}$ is plotted as a function of $k_y$ in Fig. 4e and it exhibits sign inversions at $k_y = 0$ Å$^{-1}$ and $k_y = \pm 0.5$ Å$^{-1}$ and the maxima at $k_y = \pm 0.25$ Å$^{-1}$. Similar to the DSSs, the CD in the flat band also exhibits helical-texture where opposite $k_y$ momenta have opposite signs of the CD. Notably, zero CD signal is also observed as white color in the CD-ARPES along the $k_y = 0$ Å$^{-1}$. This depicts the nodal line which was proposed to be the characteristic feature of the 2D electronic structure[36]. In particular, the CD signal of the DSSs depends on the incident photon energy assigning it to the final state effect in the photoemission process[35]. This was discussed in Ref 36 with details where they propose the non-trivial connection between the spin orbit texture and the CD signal. Thereby, the helical CD texture and the nodal line band suggest that the flat band could be also topologically non-trivial.

**Electron doping impact on the band structure:** Tuning the chemical potential in the strongly correlated systems induces a mass renormalization. This has been seen in superconductors as the mass enhancement referring to the flattening of the bands[37-38] and also exists in the CDW phase of the excitonic insulators[39]. Also, this mechanism usually coexists with a metal-insulator transition in superconductors[37]. Likewise, the dispersion of the flat band could be controlled through the electron or hole doping. However, in our case, since the flat band is already dispersionless, so the mass normalization could be expected. Namely, the flat band can gain weak dispersion upon the tuning of the $E_F$. To test this idea and to probe the possible unoccupied bands above the Fermi level, we studied the band structure of the sample under the surface deposition of potassium (K) where we found further spectroscopic anomalies in 0.3 ML $VSe_2$ / 12 QL $Bi_2Se_3$ sample (Fig. 5). The first notable observation is that K deposition induces a pair of quantum well states (QWS) shown with black arrows in Fig. 5b. This proves the electron doping induced band bending in the BVB of $Bi_2Se_3$. More interestingly, the leading edge shifts with K doping revealing an 0.08 eV insulating gap at the $E_F$ (Fig. 5a-5b). The shift is more obvious in the integrated EDCs (Fig. 5c) where sharp edges are observed at different energies before and after K deposition. This indicates that the gap closing is not due to the broadening or the formation of an additional spectral feature in the vicinity of the $E_F$. EDCs along the $k_R$ and $k_L$ momentums where the DSSs cut the $E_F$ (Fig. 5a) also exhibit the same insulating gap as shown in Fig. 5e and 5f. This shows that not only $VSe_2$ but also bare $Bi_2Se_3$ regions have an insulating gap at the $E_F$ since the size of x-ray beam spot on the sample is considerably larger than the size of individual VSe$_2$ domains. This argument is also supported by the absence of the multiple Dirac cones which are expected to be located at the bare and the VSe$_2$ covered parts of Bi$_2$Se$_3$ layers. These realizations show that the flat band could be originating from Bi$_2$Se$_3$ when modified by VSe$_2$ sublayer. Another peculiar anomaly is the difference of binding energy shifts of bulk bands, DSS, and the flat band, induced by the K deposition. While Bi$_2$Se$_3$ bulk and Dirac states exhibit the expected electron doping, the flat band shows negative electron compressibility. The flat band binding energy changes from 0.47 eV before K doping to a lower value of 0.33 eV after K doping. On the other hand, the binding energy of the DP shifts from 0.36 eV to 0.55 eV. Similar band bending towards higher binding energy is also seen in the bulk bands of Bi$_2$Se$_3$ (Fig. 5a-5b). (Note that the binding energies are given with respect to the highest occupied states in the ARPES map shown in Fig. 5b.) The existence of opposite energy shifts of the flat band and Dirac bands opens up a pathway to tune both, the binding energy of the DP as well as the flat band with respect to each other. If hole doping of Bi$_2$Se$_3$ i. e. via Ca substitution[40], was to be included, both DP and the flat band could be tuned to E$_F$ which might yield exotic transport properties. Hence, we have demonstrated the presence of a possibly unique tuning parameter for the transport properties of such low dimensional topological system.

We should also point out that the ARPES data of the K doped sample in Fig. 5b shows minute dispersion of the flat band, an apparent decrease of its binding energy when approaching the $\bar{\Gamma}$ point. However, this dispersion is too small for a precise detection but might be related to the modification of the CDW phase[39]. Our fittings of the EDCs at different k$_y$ momenta (not shown) yield the energy dispersion to be about 35 meV between k$_y$ = 0.15 Å$^{-1}$ and k$_y$ = 0.6 Å$^{-1}$ momentum points. This effect might be further enhanced by tuning the chemical potential of the system which opens an opportunity for future studies. Nevertheless, the insulating gap and the mass normalization with the electron doping, even if it is small, suggest that the flat band could be related to strong electron correlation effects.

**Discussion:** Dispersionless flat bands can be explained by several mechanisms. V 3d derived magnetic impurity bands are the first mechanism to be considered as the spectroscopic origin of the flat band[41]. However, our core level spectra and STM data do not show any signature of isolated atoms. Also, our photon energy dependent ARPES and the rather complex CD-ARPES spectra indicate that the flat band should be considered to have non-trivial origin.

Another scenario to consider would be the existence of superlattices as seen in $\sqrt{3}$ x $\sqrt{3}$ silicene superstructure by STS in which the local density of states forms the electronic Kagome lattice[42]. The interface dislocation or the strain can also flatten the original bands by introducing pseudo-magnetic field term to the Hamiltonian in Moiré superstructures[4]. However, a pronounced band flattening in this scenario requires superstructure patterns with at least a few tens of nanometers periodicity which is much larger than one observed in the present case. Furthermore, the flat bands discussed within the superlattice frameworks are dispersionless only in the BZ of the superstructure which is smaller than the BZ corresponding to the unitcell[25, 42]. This mechanism gives rise to a dispersionless flat band within a small momentum window. On the other hand, the flat band reported here does not shrink into the BZ of the Moiré or the stripe pattern superstructures and it is robust against electron doping. Therefore, although we cannot entirely exclude that the flat band electronic state in $VSe_2$ / $Bi_2Se_3$ may be related to the atomic superstructures, there are other possibilities to consider[23-25, 42].

Dispersionless flat bands with an insulating gap at the $E_F$ indicate to strong electronic correlation. This could be explained by considering the Mott insulator or charge-transfer insulating states which are generally in proximity to CDW phases[16]. Indeed, the presence of corrugated bright and dark $Bi_2Se_3$ domains in the STM image (Fig. 1e) supports the persistence of the CDW phase in agreement with earlier studies[43-44]. In the Mott phase, a strong Coulomb repulsion splits d- or f-bands into the lower (occupied) and upper (unoccupied) Hubbard bands (LHB and UHB) which are separated with an insulating gap located in the vicinity of $E_F$[16]. The insulating gap in the Mott insulator correlated with the CDW phase strongly depends on the chemical potential of the system[15]. Thus, the chemical potential could shrink the gap by shifting the LHB and UHB closer to the $E_F$ and each other. In this scenario, the Hubbard bands could exhibit opposite energy shifts than other bands upon doping[15-16]. Similarly, a large Hubbard interaction could induce an antiferromagnetic order and open an energy gap at $E_F$ in a topological honeycomb lattice[45]. Furthermore, p-type Hubbard bands with an insulating ground state were also predicted in $Bi_2Se_3$ and $Bi_2Te_2Se$ upon enhancing the Coulomb repulsion[19-21]. The similarities between the Mott insulators and the $VSe_2$ / $Bi_2Se_3$ could suggest that the flat band in our sample could be formed by the strong correlation effects.

On the other hand, up to the date, the surface electronic structure of TIs has been studied by ARPES under the deposition of various magnetic and non-magnetic elements, and none of the approaches revealed the flat band spectral feature[46]. This controversy could be related to the location of the dopants in the crystal. The elemental dopants occupy the vdW gap or the sublattice sites in the bulk[47], while $VSe_2$ domains are located on the surface of $Bi_2Se_3$. Therefore, the magnetic properties of the $VSe_2$ could also play a vital role in the formation of the distinct electronic states on the surface of $Bi_2Se_3$ without modifying its bulk band topology[48]. Nevertheless, the details of the flat band electronic structure with a realistic scenario are more complicated since non-trivial band topology and the strong spin-orbit coupling are also expected to play a role in the evolution of the electronic structure. The complete understanding of the physics behind the flat band and the insulating gap requires a broad and elaborate theoretical and experimental efforts. As a first example of the dispersionless electronic excitation in a topologically non-trivial band structure, our results could open a new pathway in the critical field of experimental realization and control of novel quantum effects and new states of the matters.

**Methods:**

**Material preparation:** Molecular beam epitaxial growth (MBE) technique was employed to grow $VSe_2$ / $Bi_2Se_3$ and $Bi_2Se_3$ samples in a custom ultra-high vacuum system located at the ESM beamline of NSLS-II. 5N Se and Bi sources were evaporated from the ceramic crucibles while the e-beam evaporation method was used for V (99.8% purity) source. All samples were grown on $Al_2O_3(0001)$ substrates at 255 °C. Before the growth, the substrates were first degassed at 550 °C for three hours and flashed at 850 °C for 5 min. Sample thicknesses were estimated within a 15 % error bar by using a quartz thickness monitor and x-ray photoemission spectroscopy. Samples for ARPES and μLEED experiments were capped with 20 nm amorphous Se film before being removed from the MBE chamber.

**XPS:** Core-levels were recorded with 1486.7 eV non-monochromated x-ray source at room temperature with a vacuum interlocked MBE-photoemission system at the surface science laboratory of the Department of Physics of the University of Connecticut. Core-level binding energies were determined with an error of 0.1 eV.

**ARPES:** ARPES experiments were performed at 21-ID-1 ESM beamline of National Synchrotron Light Source II (NSLS-II) by using a DA30 Scienta electron spectrometer. The pressure in the photoemission chamber was $1 \times 10^{-10}$ Torr and samples were kept at 15 K during the experiment by a closed-cycle He cryostat. The energy resolution in the ARPES experiments was better than 15meV with a spot size of ~20 µm. Before the ARPES experiments, samples were annealed at 220 ºC for 30 minutes to remove the Se capping layer. The angle between the light and the surface normal of the sample is 55° at the normal emission during the ARPES experiments. The films were grounded with a tantalum clip. A part of the ARPES experiments was conducted at the linear undulator beamline at the Hiroshima Synchrotron Radiation Center (HiSOR) BL-1 [S1]. Photon energy is converted to $k_z$ momentum space by using the free electron final state approximation $\hbar k_z = \sqrt{2m_e(E_{kin}\cos^2\theta + V_o)}$ where $m_e$ is the free electron mass, $E_{kin}$ is the kinetic energy of a photoelectron, and $V_o$ is the inner potential which is 11.8 eV for $Bi_2Se_3$ [S2].

**TEM and STM:** HAADF-STEM images were acquired with Hitachi HD2700C dedicated STEM with a probe Cs corrector operating at 200 kV at room temperature. Samples were prepared using the in-situ lift-out method on the FEI Helios 600 Nanolab dual-beam FIB. Final milling was completed at 2 keV. Scanning tunneling microscope (Omicron VT- STM -XA 650 ) experiments were performed in an ultrahigh vacuum (UHV) system with a base pressure of 2 × 10−10 Torr at room temperature. All the STM images were observed in constant current mode using Pt/ Ir tips. All bias values in the text refer to the bias applied to the sample. The STM images were analyzed using Gwyddion-2.55 software package. HAADF-STEM and STM experiments were conducted at the Center for Functional Nanomaterials, Brookhaven National Laboratory. Samples for STM were transferred with a vacuum suitcase.

**LEED:** µLEED experiment was performed at XPEEM/LEEM endstation of the ESM beamline (21-ID-2).

**Acknowledgment:** This research used ESM (21-ID-1, 21-ID-2) beamline of the National Synchrotron Light Source II, a U.S. Department of Energy (DOE) Office of Science User Facility operated for the DOE Office of Science by Brookhaven National Laboratory under Contract No. DE-SC0012704. This work also used the resources of the Center for Functional Nanomaterials, Brookhaven National Laboratory, which is supported by the U.S. Department of Energy, Office of Basic Energy Sciences, under Contract No. DE-SC0012704. ARPES experiments in Hiroshima were performed with the approval of program advisory committee of HISOR) Proposal No. 19BG041). Author T. Y. would like to thank Prof. A. V. Balatsky for useful discussions.

**Author Contributions:** T.Y conceived and designed the experiments. T.Y prepared the samples and performed the photoemission experiments with K. K., E. V, and B. S's help. E. F. S. performed the ARPES experiments at HISOR. X. T. conducted the STM measurements. Z.D. and J. T. S. performed µLEED measurements. S. H. and K. K. performed HAADF-STEM experiments. T. Y analyzed the experimental results and wrote the manuscript with contribution from E. F.S., B. S., K. K., and E. V.

**Competing interests:** The authors declare that they have no competing interests.

**Figures:**

**Fig. 1 | Structural characterization. a,** Top and side views of $Bi_2Se_3$ and $VSe_2$ crystal structures. Hexagonal BZs with the high-symmetry points are given in the lower part of **a**. **b,** Core-level photoemission spectra of Bi 5d and Se 3d obtained from 12 QL $Bi_2Se_3$, 0.3 ML $VSe_2$ / 12 QL $Bi_2Se_3$, and 3 ML $VSe_2$ / 12 QL $Bi_2Se_3$ samples. The inset in **b** shows the Se 3d peaks for pristine and 3 ML $VSe_2$ covered $Bi_2Se_3$ after scaling peaks to the same height for visual comparison. **c,** V 2p core levels for 0.3 ML $VSe_2$ / 12 QL $Bi_2Se_3$ and 3 ML $VSe_2$ / 12 QL $Bi_2Se_3$ samples. The atomic stoichiometry of Se to V is computed to be 2 by using the peak areas and photoionization cross-sections. **d,** HAADF-STEM cross-section image of 3 ML $VSe_2$ / 12 QL $Bi_2Se_3$ heterostructure. The color contrast in **d** is correlated with the atomic number (Z-contrast). **e, f,** Room temperature STM images of $Bi_2Se_3$ (at sample bias 100 mV, set point 1 nA) and $VSe_2$ (at sample bias 80 mV, set point 1 nA) surfaces, respectively obtained from 0.3 ML $VSe_2$ / 12 QL $Bi_2Se_3$. Yellow parallelogram in **f** represents the Moiré unit cell.

**Fig. 2 | Emergence of the flat band revealed by ARPES. a-b,** Experimental electronic structures of 12 QL $Bi_2Se_3$ and 0.3 ML $VSe_2$ / 12 QL $Bi_2Se_3$ samples, respectively. Spectra were collected with 50 eV linear horizontal polarized lights. BVB refers to the bulk valance band of $Bi_2Se_3$. **c,** Momentum integrated EDCs of 0.3 ML $VSe_2$ / 12 QL $Bi_2Se_3$ acquired with 110 eV photon energy along the $\bar{\Gamma}$-$\bar{M}$ direction in the BZ. **d-e,** Constant energy counters at the $E_F$ and the $E_{FB}$, respectively for 0.3 ML $VSe_2$ / 12 QL $Bi_2Se_3$. Red and black hexagons in **e** correspond to the BZ of $VSe_2$ and $Bi_2Se_3$, respectively. In-plane lattice parameters of 4.14 Å for $Bi_2Se_3$ and 3.356 Å for $VSe_2$ were employed to compute the BZs[27, 31].

**Fig. 3 | Photon energy-dependent ARPES experiment. a-b,** $k_y$ - $k_z$ dispersions at the $E_F$ and the $E_{FB}$, respectively. Dashed red lines in **a** mark the DSSs. **c,** MDCs at different $k_z$ points **d-e,** Binding energy versus $k_z$ maps at $k_y = \pm 0.3$ Å, respectively. Dashed cyan colored lines in **d** and **e** represent the modulation of the flat band along the $k_z$ direction. **f,** EDCs at various $k_z$ points to study the spectral shape of the flat band. ARPES maps for the plots were conducted along the $\bar{\Gamma}$-$\bar{M}$ direction in the BZ.

**Fig. 4 | CD-ARPES. a-b,** ARPES maps of 0.3 ML $VSe_2$ / 12 QL $Bi_2Se_3$ sample recorded with RCP and LCP lights, respectively. **c,** Computed CD-ARPES. **d-e,** CDs as a function of $k_y$ at 0.1 eV binding energy and the $E_{FB}$, respectively.

**Fig. 5 | Electron doping impact on the band structure. a-b,** ARPES maps of 0.3 ML $VSe_2$ / 12 QL $Bi_2Se_3$ before (bare) and after K-doping, respectively. Spectra were obtained with 50 eV photons at 15 K. **c-f** present integrated EDC within $k_y = \pm 0.6$ Å and EDCs along the $k_0$, $k_R$, and $k_L$ momentums as marked in **a**. In **a-b**, dashed lines show the energy shifts of the different bands as specified in the figures and the yellow solid line indicates the leading edge. In **c-f**, green, pink, black, brown vertical lines represent the binding energy of the flat band, leading edge, bulk bands, and the DP, respectively. The arrows in **c-f** point the direction of the energy shifts.

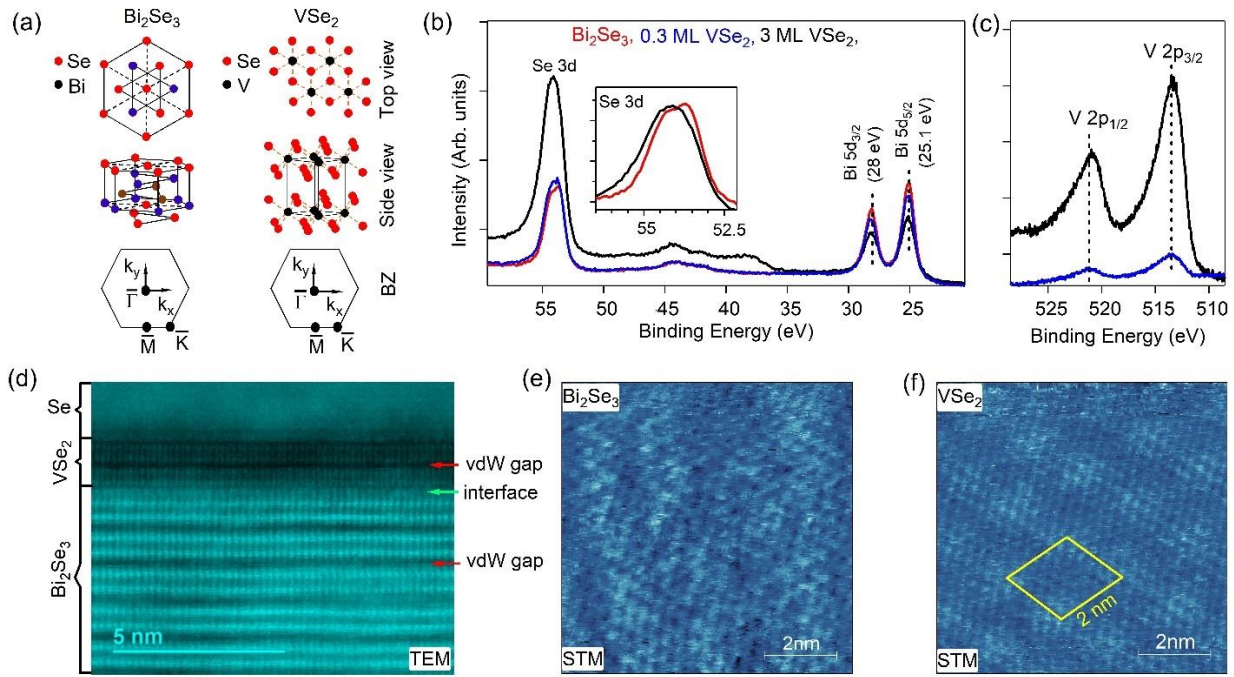

**Fig. 1**

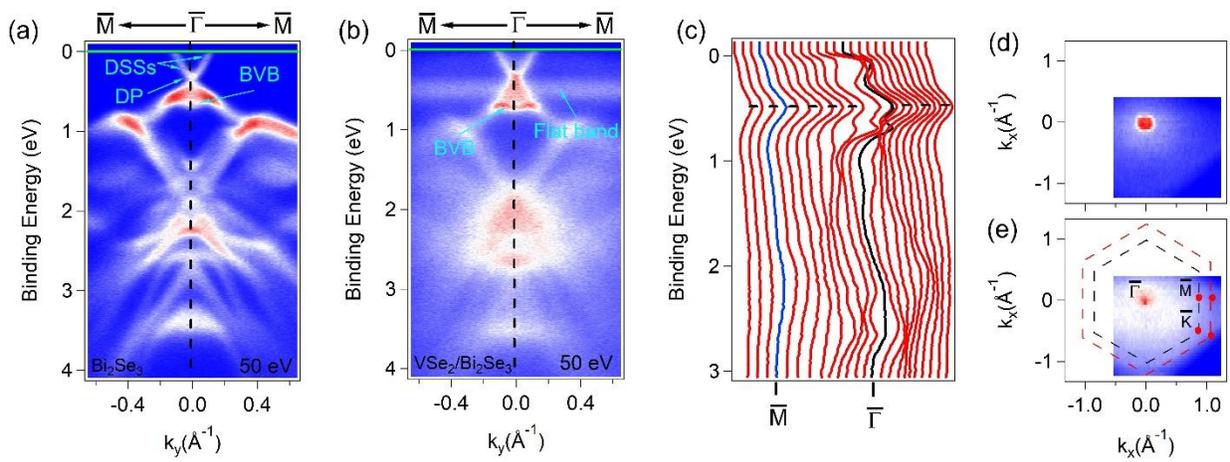

**Fig. 2**

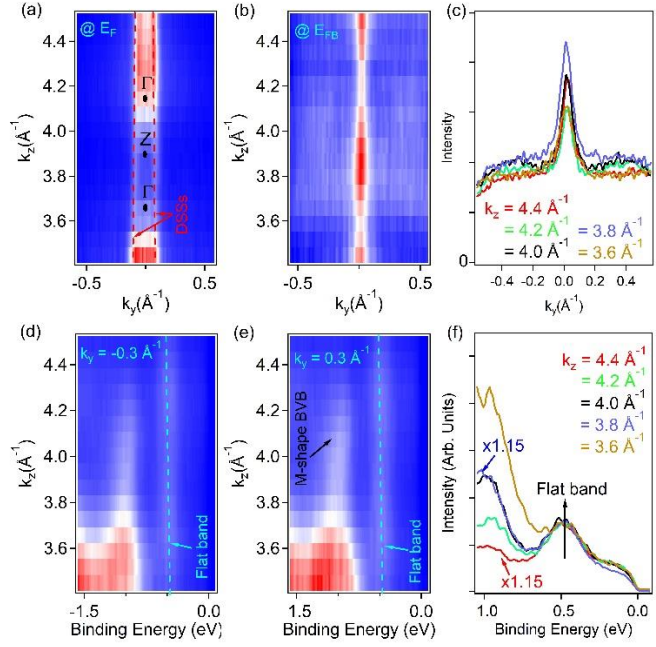

**Fig. 3**

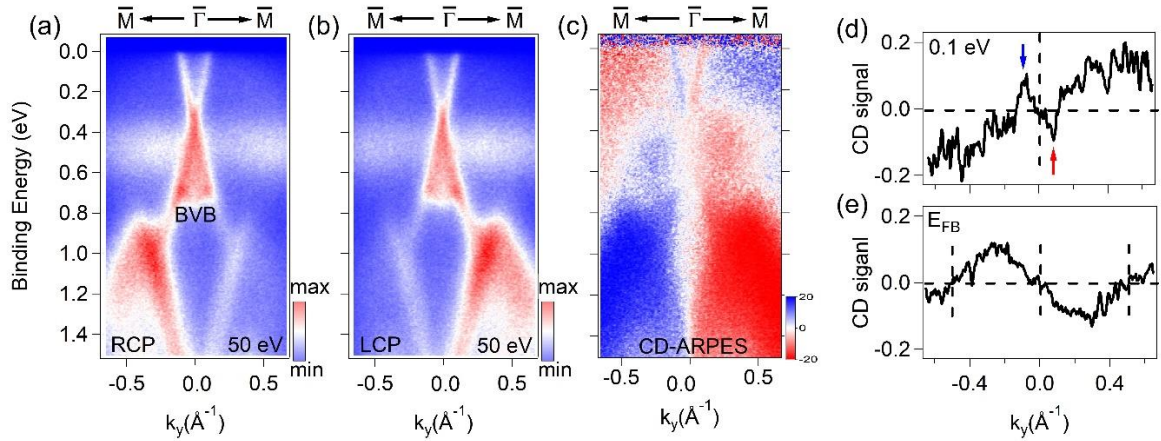

**Fig. 4**

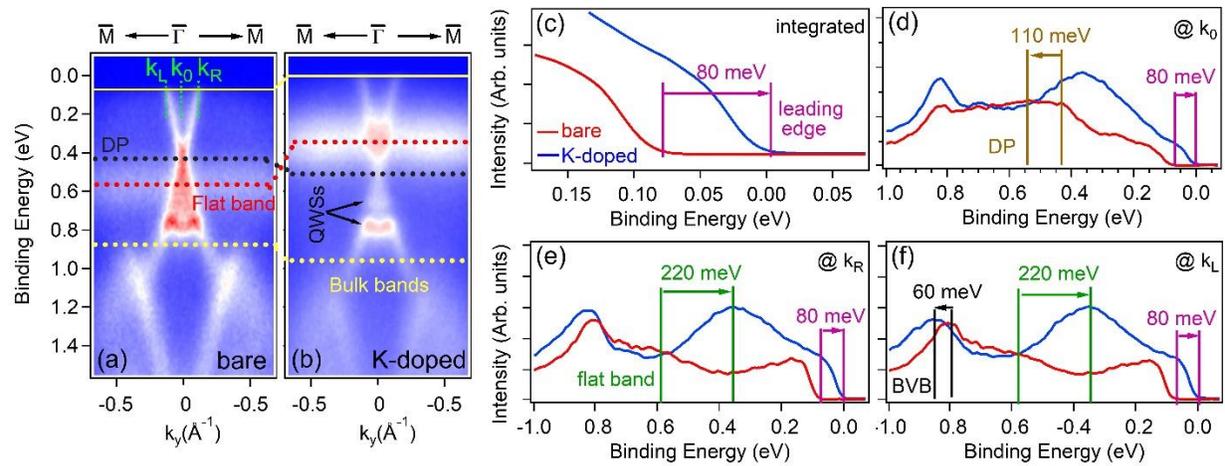

**Fig. 5**